\documentclass[prl,twocolumn,showpacs,floatfix]{revtex4-1}

\usepackage{graphicx}
\usepackage{array}
\usepackage{amsmath,amsfonts,amssymb}
\usepackage[ansinew]{inputenc}
\usepackage{color}

\newcommand{\buck}{C$_{60}$}
\newcommand{\Figref}[1]{Fig.~\ref{#1}}

\begin{document}
\title{Imaging isodensity contours of molecular states with STM}
\author
{Ga\"el Reecht,$^{1,2}$ Benjamin Heinrich,$^2$ Herv\'e Bulou,$^1$ Fabrice Scheurer,$^1$ Laurent Limot,$^1$ Guillaume Schull,$^{1}$ }

\altaffiliation{guillaume.schull@ipcms.unistra.fr}
\affiliation{$^1$ Universit\'e de Strasbourg, CNRS, IPCMS, UMR 7504, F-67000 Strasbourg, France}
\affiliation{$^2$ Fachbereich Physik, Freie Universit\"at Berlin, 14195 Berlin, Germany}

\begin{abstract}
We present an improved way for imaging the local density of states with a scanning tunneling microscope, which consists in mapping the surface topography while keeping the differential conductance (d$I$/d$V$) constant. When archetypical C$_{60}$ molecules on Cu(111) are imaged with this method, these so-called iso-d$I$/d$V$ maps are in excellent agreement with theoretical simulations of the isodensity contours of the molecular orbitals.  A direct visualization and unambiguous identification of superatomic C$_{60}$ orbitals and their hybridization is then possible.     
\end{abstract}

\date{\today}


\maketitle

The ability of the scanning tunneling microscope (STM) to image and address conductive surfaces at the atomic-scale is the main reason of its impressive success over the last three decades. Of appealing interest is the possibility of probing the local density of states (DOS) of metallic or organic nanostructures adsorbed on surfaces. If the energy distribution of the local DOS is usually directly inferred from differential conductance (d$I$/d$V$) spectra, a strict comparison requires a renormalization of the experimental data \cite{Stroscio1986,Ukraintsev1996,Wagner2007,Koslowski2007,Passoni2009}. Measuring the spatial distribution of the DOS is even more demanding. Maps of the DOS are usually obtained by recording the d$I$/d$V$ at a given target voltage ($V$) while keeping the tunneling current ($I$) constant \cite{Hasegawa1993,Kedong2003,Nilius2005,Temirov2006,Feng2008,Schull2008,Schull2009,Auwaerter2010,Mullegger2011,Ruffieux2012,Fahrendorf2013,Nacci2015,Pham2016}. However, these so-called constant-current d$I$/d$V$ maps suffer from the fact that the tip--sample distance ($z$) varies during the scan and therefore do not properly reflect the local DOS of the probed nanostructure. This well-known limitation has been evidenced for surface--confined electronic states \cite{Hoermandinger1994, Li1997}, adatoms \cite{Ziegler2009} or molecules adsorbed on surface \cite{Lu2003,Krenner2013,Zhang2015}. To extract reliable information from these maps, a time-consuming image treatment is required after acquisition \cite{Li1997,Lu2003,Ziegler2009,Krenner2013,Zhang2015}.

An interesting alternative consists in recording the d$I$/d$V$ while scanning the surface with an open feedback loop \cite{Ziegler2009,Heinrich2010,Krenner2013,Reecht2013,Zhang2015,Hajo2015}. In the case of a flat sample, the obtained image -- the so-called constant-height d$I$/d$V$ map -- is an accurate representation of the local DOS. However, this method is limited in the case of corrugated objects, because the effective tip--sample distance varies as a function of the $(x,y)$ position of the tip. Moreover, since the data are acquired with a disabled feedback loop, this method requires small thermal drift during the acquisition of the conductance map, limiting the field of application to cryogenic measurements which provide sufficient thermal stability.

In this letter, we propose a different experimental approach that enables a direct visualization of DOS isosurfaces, independently of the corrugation of the sample, and that may be implemented at all working temperatures. This imaging technique consists in acquiring iso-d$I$/d$V$ maps, in other words in scanning the tip across the surface while keeping the d$I$/d$V$ signal constant (instead of the current) and recording changes in $z$. Through a combination of experiments with  density functional theory (DFT) calculations, we show that the iso-d$I$/d$V$ maps of non-planar \buck\ molecules on a Cu(111) surface closely reflect theoretical representations of the molecule DOS isosurface unlike standard d$I$/d$V$ maps, \textit{i.e.}, constant-current and constant-height, that exhibit misleading patterns for some of the \buck\ orbitals~\cite{Lu2003}. We use this new method to provide a fresh insight into the recently reported superatomic orbitals of \buck\ monomers and dimers~\cite{Feng2008}.  

\begin{figure}
  \includegraphics[width=85mm,clip=]{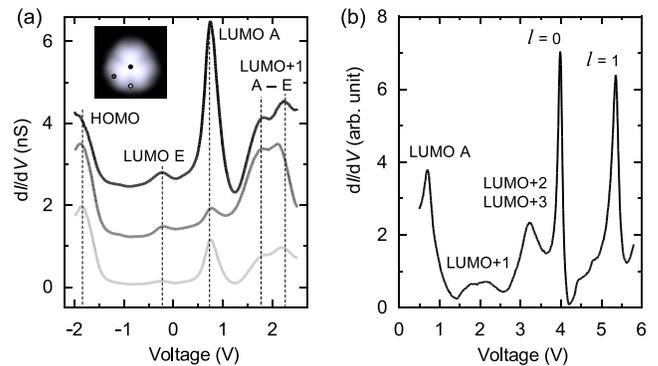}
  \caption{(a)  Constant-height d$I$/d$V$ spectra acquired for tip positions marked by the dots in the STM image in inset (2.3 $\times$ 2.3 nm$^2$, $V$ = 2.2 V). (b) Constant-current d$I$/d$V$ spectrum for the tip positioned above the center of \buck\ ($I$ = 1 nA).}
  \label{fig1}
\end{figure}

The experiments were performed with a STM operated at 4.6\,K in UHV. Electrochemically etched W tips and Cu(111) samples were prepared by successive cycles of Ar$^+$ bombardment and annealing. The \buck\ molecules were sublimated from an evaporator onto the cold ($\approx$ 5\,K) sample. The d$I$/d$V$ spectra were recorded via a lock-in amplifier by applying an AC bias of 7 kHz modulation frequency and of 10 mV rms amplitude (50 mV rms for the maps). Except for \Figref{fig1}b, the d$I$/d$V$  spectra were acquired with a disabled feedback loop. To permit an unbiased comparison between the different imaging methods, the experimental maps were not processed, except for the pseudo-3D representations in Fig.~\ref{fig4}, which were treated with a smoothing algorithm. All the images were prepared with the Nanotec Electronica WSxM software \cite{Horcas2007}. Details of the DFT calculations are provided in the Supplemental Material \cite{SI}.

To start, we briefly describe the working principle of an iso-d$I$/d$V$ map. At low temperature and assuming a constant tip DOS, the tunnel current can be expressed as $I(z,V)\propto \int_{0}^{eV} \rho_s(E)\:T(z, V, E)\:\mathrm{d}E$ \cite{Simmons1963} where $\rho_s$ is the surface DOS and $E$ the energy. Within the Wentzel-Kramer-Brillouin approximation, the transmission factor reads  $T(z, V, E) \propto \mathrm{exp}(-\alpha z \sqrt{\phi + eV/2-E})$ where $\phi$ is the local barrier height and $\alpha = 2\sqrt{2m}/\hbar$ ($m$: free electron mass, $\hbar$: reduced Planck constant). The derivative of the current with respect to $V$ reads 

\begin{eqnarray}
\label{eq:di/dv}
 \frac{\mathrm{d} I(z,V)}{\mathrm{d}V} \propto e\rho_s(eV)T(z, V, eV)\nonumber\\
 + \int_{0}^{eV} \frac{\partial T(z, V, E)}{\partial V}\:\rho_s(E)\:\mathrm{d}E \nonumber\\
 + \int_{0}^{eV}\:\frac{\mathrm{d}z}{\mathrm{d}V}\:\frac{\partial T(z, V, E)}{\partial z}\:\rho_s(E)\:\mathrm{d}E.
\end{eqnarray} 
In the following, we disregard the third term of Eq.~(\ref{eq:di/dv}) as the lock-in modulation frequency of the AC bias is purposely chosen to be high compared to the time constant of the feedback loop ($dz/dV=0$). We then find the usual expression~\cite{Koslowski2007,Ziegler2009}
\begin{equation}
\label{eq:simple ro}
\rho_s(eV) \propto \frac{1}{eT(z, V, eV)} \left[\frac{\mathrm{d} I(z,V)}{\mathrm{d} V}+\frac{z}{4\alpha\sqrt{\phi}}I(z,V)\right],
\nonumber
\end{equation} 
where the second term can be neglected hereafter as discussed in Supplemental Material \cite{SI}. Since the d$I(z,V)$/d$V$ signal is kept constant during the image acquisition, we then have at a given ($x,y$) position of the surface
\begin{equation}
\label{z}
z(x, y, V) \propto \ln{\rho_s(x, y, eV)}.
\end{equation} 
In other words, Eq.~(\ref{z}) shows that it is possible in this way to directly measure the isodensity contours of the sample DOS at a given energy $eV$. We stress that the level of approximation employed here is the same generally used when discussing standard d$I$/d$V$ maps.

Next, to validate our imaging technique, we carry out a comparative study between standard d$I$/d$V$ maps and iso-d$I$/d$V$ maps by focusing our attention onto isolated \buck\ molecules on Cu(111). In the inset of \Figref{fig1}a we present a standard STM image of a \buck\, where the characteristic threefold-symmetric shape of the molecule can be recognized, indicating that it is adsorbed with a hexagon oriented towards the tip \cite{Larsson2008,Heinrich2011,Shin2014}. The d$I$/d$V$ spectra (Figs.~\ref{fig1}a) acquired for different positions of the STM tip (dots in the STM image in the inset of \Figref{fig1}a) reveal a variety of molecular resonances. Following Ref.~\cite{Silien2004}, we identify the highest occupied molecular orbital (HOMO) and the two lowest unoccupied molecular orbitals (LUMO and LUMO+1). Because of the strong interaction with the Cu(111) substrate, the LUMO and LUMO+1 orbitals are split (A and E components). The spectrum in \Figref{fig1}b reveals further resonances at higher energy.  We assign the broad peak at $3.2$\,V to the superposition of LUMO+2 and LUMO+3 states. The DFT calculations for the isolated, unstrained molecule reveal in fact that these states are separated by only 200\,meV \cite{SI}, which is below the lifetime broadening of the molecular states ($\approx 300\,$meV). Additionally, two sharp resonances appear at $4$ V and $5.2$ V that correspond to superatomic $l$ = 0 and $l$ = 1 states~\cite{Feng2008}, which can also be understood in the framework of whispering gallery modes \cite{Reecht2016}.

\begin{figure*}
  \includegraphics[width=150mm,clip=]{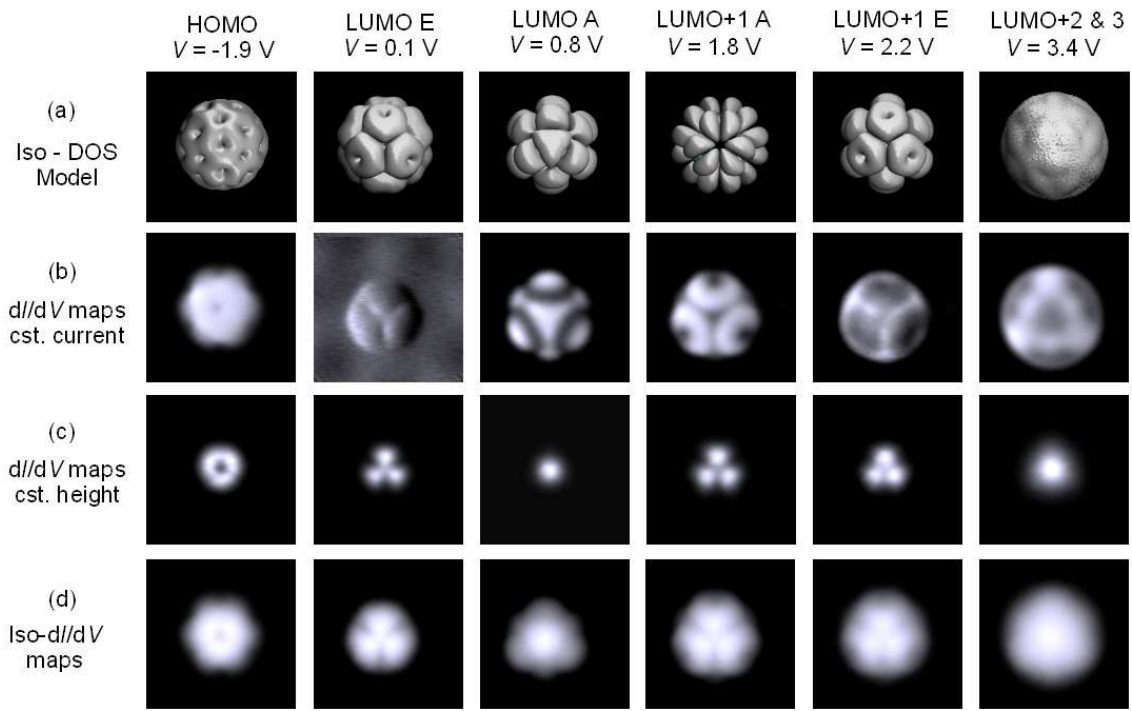}
\caption{(a) Isosurface contours of the \buck\ orbitals computed with DFT (10$^{-5}$ e/bohr$^3$). (b) Constant-current d$I$/d$V$ maps ($I$ = 1 nA), (c) constant-height d$I$/d$V$ maps (feedback opened at $I$ = 1 nA for a tip located on the center of the \buck) (d) iso-d$I$/d$V$ maps (d$I$/d$V$ = 0.7 nS, corresponding to $I \leq$ 1 nA) of an individual \buck\ acquired at the biases identified in the spectra of \Figref{fig1}. The images have the same size ($3.1\times 3.1$ nm$^2$).}
\label{fig3} 
\end{figure*} 

Knowing the molecular orbitals energies, we can now measure their spatial distribution.  To start, we display in Fig.~\ref{fig3}a the computed DOS isosurfaces for each molecular orbitals. These gas phase DFT calculations take into account the degeneracy lifting caused by the interaction with the surface~\cite{SI}. While the HOMO orbital does not split on the surface, the threefold degeneracy of the LUMO and of the LUMO+1 orbitals is partially lifted upon adsorption \cite{Lu2003,Silien2004}. Following simple symmetry arguments~\cite{Hands2010}, we decomposed the computed LUMO and LUMO+1 orbitals into their A and E components. Finally, since the LUMO+2 and LUMO+3 orbitals cannot be distinguished in the d$I$/d$V$ spectra, we consider only the sum over the LUMO+2 and LUMO+3 isosurfaces.

\begin{figure*}
  \includegraphics[width=150mm,clip=]{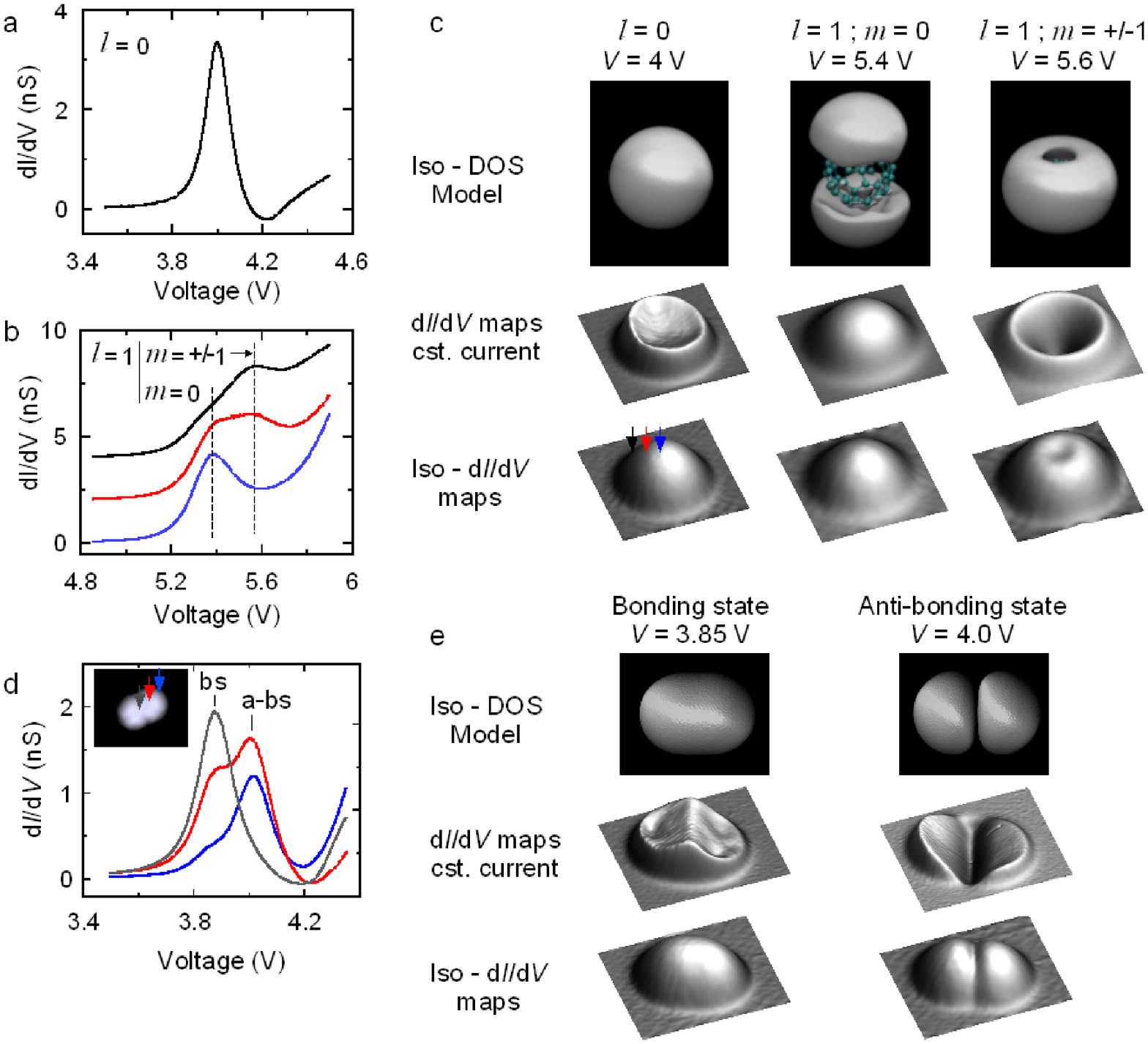}
\caption{d$I$/d$V$ spectra of (a) the $l$ = 0 and (b) the $l$ = 1 resonances of a single \buck\ molecule on Cu(111). The spectra in (b) were acquired for three different lateral positions of the tip with respect to the \buck\ molecule as indicated by the colored arrows in panel (c). (c) Computed DOS isosurfaces (top row, 10$^{-5}$ e/bohr$^3$), constant-current d$I$/d$V$ maps (middle row, $I$ = 1 nA for $l$= 0 and $I$ = 0.3 nA for $l$ = 1 maps) and iso-d$I$/d$V$ maps (bottom row, d$I$/d$V$ = 0.8 nS) of an individual \buck\ acquired at the biases identified in the spectra in (a) and (b) (size of the theoretical images: 4.0 $\times$ 5.4 nm$^2$ and of the experimental images: 4 $\times$ 4 nm$^2$). (d) d$I$/d$V$ spectra acquired for three different lateral positions of the tip with respect to a \buck\ dimer as indicated by the colored arrows in the STM image in the inset of panel (d). (e) Computed DOS isosurfaces (top row, 10$^{-5}$ e/bohr$^3$), constant-current d$I$/d$V$ maps (middle row, $I$ = 1 nA) and iso-d$I$/d$V$ maps (bottom row, d$I$/d$V$ = 0.2 nS) for a \buck\ dimer acquired at the biases identified in the spectra in (d) (The images have the same size: 5.9 $\times$ 4.9 nm$^2$).}
\label{fig4} 
\end{figure*}

Experimentally, the spatial distributions of the above orbitals have been probed following three different approaches: with constant-current d$I$/d$V$ maps (\Figref{fig3}b), constant-height d$I$/d$V$ maps (\Figref{fig3}c), and iso-d$I$/d$V$ maps (\Figref{fig3}d). The agreement between these maps and the calculated DOS isosurfaces will be discuss separately. The constant-current d$I$/d$V$ maps (\Figref{fig3}b) agree well with the simulation for the HOMO, the LUMO A and the LUMO+1 A. At the energy of the LUMO E, the signal is very low and the pattern is asymmetric. For the E component of the LUMO+1, the constant-current d$I$/d$V$ map reveals a pattern of inverted contrast compared to the calculation. This behavior, discussed in detail by Lu \textit{et al.}, is due to variations of the tip-sample distance during the data acquisition \cite{Lu2003}. Possible artifacts and the resulting inaccuracy of this method will be further discussed in the next section.

The d$I$/d$V$ maps (\Figref{fig3}c) recorded at a constant tip height respect the symmetry of the computed isosurfaces. However, the level of detail is reduced compared to the other approaches. More precisely, only the top part of the molecule is imaged. This is a direct consequence of the sphericity of \buck\ showing that, for a corrugated surface, constant-height measurements produce limited results.

In contrast, the iso-d$I$/d$V$ maps in Fig.~\ref{fig3}d provide a correct representation of the spatial variation of all states. Using this method, the STM tip directly follows the isosurface DOS of a molecular state. This largely facilitates the identification of the molecular orbitals and prevents possible misinterpretations. As an example, here we can clearly identify the LUMO A at $V=0.8$\,V, an orbital whose assignement was always uncertain in previous measurements \cite{Silien2004} because of its similarity with the LUMO+1 E in constant-current d$I$/d$V$ maps, and whose identification is not possible in constant-height maps where it only appears as a bright protrusion. 

After this first proof of principle, we now use the iso-d$I$/d$V$ maps to unveil the spatial distribution and composition of the resonances assigned to the $l$ = 0 and $l$ = 1 superatomic states. The corresponding d$I$/d$V$ spectra of these states are displayed in Figs.~\ref{fig4}a and \ref{fig4}b, respectively. The computed DOS isosurface for the $l$ = 0 state (\Figref{fig4}c) reveals a uniform sphere corresponding to a state fully delocalized over the \buck\ cage. While the constant-current d$I$/d$V$ map exhibits a hole-like structure that would suggest a hybridization between the $l$ = 0 and the $l$ = 1 orbitals \cite{Feng2008}, the iso-d$I$/d$V$ map is instead in perfect agreement with the theoretical predictions. 

The $l$ = 1 superatomic state is threefold degenerate ($m =-1,0,1$) in vacuum, but the d$I$/d$V$ spectra acquired at different positions above the molecule on copper (see arrows in \Figref{fig4}c) reveal instead a split resonance for this state (\Figref{fig4}b). This suggests that the interaction with the surface lifts the degeneracy among the $m$ states. Our DFT computations indicate that the state at $5.4$~V is associated to $m=0$ and the state at $5.6$~V to $m=\pm1$. In \Figref{fig4}c, we present the d$I$/d$V$ maps of these states along with their computed DOS isosurfaces. Again, it can be remarked that the iso-d$I$/d$V$ maps are in good agreement with calculations. The constant-current d$I$/d$V$ map, instead, unsatisfactorily reproduce the $m=\pm1$ contribution: the ring diameter is too large, the signal fall-of towards the center steep, and the signal at the center of the molecule is lower than the signal on the substrate.

The discrepancy between constant-current and iso-d$I$/d$V$ maps is even more striking for a \buck\ dimer (Figs.~\ref{fig4}d and~\ref{fig4}e). Here, the hybridization between the $l$ = 0 superatomic states of the molecules leads to a splitting of the orbital into a bonding state (bs) and an anti-bonding state (a-bs) (Figs.~\ref{fig4}d). Contrary to constant-current d$I$/d$V$ maps, the bounding and anti-bounding states can be readily visualized with iso-d$I$/d$V$ maps, their pattern being self-explanatory and in perfect agreement with simulations.  

To summarize, we presented a simple way to accurately map with STM the spatial variation of the DOS, which is well-suited for non-planar molecules and artificial nanostructures. Because the feedback loop is enabled during the data acquisition, our method is applicable to corrugated surfaces and in the presence of thermal drift. By imaging in this way individual \buck\ molecules on Cu(111) and comparing the results to DFT calculations, we could unambiguously identify the different resonances in the $dI/dV$ spectra, in particular the A component of the LUMO orbital.  Furthermore, we were able to correctly visualize the spatial distribution of the superatomic states in the \buck\ monomer, as well as their hybridization in the dimer case. The iso-d$I$/d$V$ maps are therefore an excellent error-free alternative to commonly DOS mapping techniques employed with STM.

\noindent
We thank V. Speisser, J.-G. Faullumel and M. Romeo for technical support. The Agence National de la Recherche (Grant No. ANR-14-CE26-0016-01, ANR-15-CE09-0017, ANR-13-BS10-0016, ANR-10-LABX-0026 CSC, ANR-11-LABX-0058 NIE) and the International Center for Frontier Research in Chemistry are acknowledged for financial support. This work was performed using HPC resources from GENCI-IDRIS (Grant No. 2016097459). BWH gratefully acknowledges funding by the Deutsche Forschungsgemeinschaft through Grant HE7368/2.

\end{document}